\def\draftversion{false}
\RequirePackage{ifthen}
\ifthenelse{\equal{\draftversion}{true}}{
  \documentclass[floatfix,galley,prb]{revtex4}
}{
  \documentclass[floatfix, twocolumn, prb]{revtex4}
}

\usepackage{color}
\usepackage{multirow}
\usepackage{epsfig}
\usepackage{graphicx}
\usepackage{amsmath}
\usepackage{amssymb}
\usepackage{bm}
\usepackage{dcolumn}
\usepackage{braket}
\usepackage{bbold}
\usepackage{sidecap}
\usepackage{float}
\usepackage{hyperref}
\hypersetup{hidelinks}
\usepackage[normalem]{ulem}
\usepackage{calrsfs}
\DeclareMathAlphabet{\pazocal}{OMS}{zplm}{m}{n}
\newcommand{\Lt}{\pazocal{T}}
\newcommand{\Lp}{\pazocal{P}}

\begin{document}

\title{Magnetic i-MXene: a new class of multifunctional two-dimensional materials}
\author{Qiang~Gao}
\author{Hongbin~Zhang}
\email[corresp.\ author: ]{hzhang@tmm.tu-darmstadt.de}
\affiliation{Institute of Materials Science, Technische Universit$\ddot{a}$t Darmstadt, 64287 Darmstadt, Germany}

\date{\today}

\begin{abstract}
  Based on density functional theory calculations, we investigated the two-dimensional in-plane ordered MXene (i-MXenes), focusing particularly on the magnetic properties.
It is observed that robust two-dimensional magnetism can be achieved by alloying nonmagnetic MXene with magnetic transition metal atoms. 
Moreover, both the magnetic ground states and the magnetocrystalline anisotropy energy of the i-MXenes can be effectively manipulated by strain, indicating strong piezomagnetic effect.  
Further studies on the transport properties reveal that i-MXenes provide an interesting playground to realize large thermoelectric response, antiferromagnetic topological insulator, and spin-gapless semiconductors.
Thus, i-MXenes are a new class of multifunctional two-dimensional magnetic materials which are promising for future spintronic applications.

\end{abstract}

\maketitle

\def\scr{\scriptsize}
\ifthenelse{\equal{\draftversion}{true}}{
  \marginparwidth 2.7in
  \marginparsep 0.5in
  \newcounter{comm} 
  \def\commnext{\stepcounter{comm}}
  \def\commtext{{\bf\color{blue}[\arabic{comm}]}}
  \def\commmar{{\bf\color{blue}[\arabic{comm}]}}
  \def\hzm#1{\commnext\marginpar{\small HZ\commmar: #1}\commtext}
  \def\qgm#1{\commnext\marginpar{\small QG\commmar: #1}\commtext}
}{
  \def\hzm#1{}
  \def\qgm#1{}
}

\def\Red#1{\textcolor{red}{#1}}
\def\Blue#1{\textcolor{blue}{[#1]}}
\def\Magenta#1{\textcolor{magenta}{#1}}
\section{Introduction}
\label{intro}


The MAX compounds with a general chemical formula of M$_{n+1}$AX$_n$ (M: early transition metal; A: main group element; X: C or N; n: integer up to 6)
have drawn intensive attention recently, due to potential applications as structural, electrical, and tribological materials.~\cite{sun_progress_2011} 
The corresponding two-dimensional (2D) nanosheets M$_{n+1}$X$_n$, dubbed as MXene, can be obtained by etching the A elements away, {\it e.g.}, Ti$_3$C$_2$ from Ti$_3$AlC$_2$.~\cite{naguib_two-dimensional_2011}
 This defines a new class of 2D materials beyond graphene,~\cite{naguib_25th_2014}
with a number of materials reported showing multi-functionality covering 
energy, spintronic, nano-electronic, and topological applications.~\cite{anasori_2d_2017, gogotsi_rise_2019}  
For instance, theoretical calculations predicted that Ti$_3$C$_2$ is a high performance anode material for lithium ion batteries,~\cite{takacs_solar_2012} 
which was confirmed later by experiments.~\cite{lukatskaya_cation_2013}
Although most MXenes are metallic due to the partially occupied $d$-shells of the transition metal atoms, 
functionalization can be applied to open up a finite band gap thus further tailor their properties. 
For instance,  passivated MXene Mo$_2$MC$_2$ (M = Ti, Zr and Hf) with the O$_2$ group are found to be robust quantum spin Hall insulators 
with band gaps from 0.1 eV to 0.2 eV.~\cite{si_large-gap_2016} 
Particularly, the magnetic properties of MXenes deserve further investigation, driven by the discovery of 2D magnetic monolayers 
such as CrI$_3$ and Cr$_2$Ge$_2$Te$_6$.~\cite{huang_electrical_2018,huang_layer-dependent_2017,gong_discovery_2017} 
Gao {\it et al.}~\cite{gao_monolayer_2016} predicted that Ti$_2$C and Ti$_2$N are nearly half-metals, 
which can be tuned into spin-gapless semiconductor (SGS) under biaxial strain.  
Theoretical calculations also demonstrated that Mn$_2$N with functional groups O, OH and F can be half-metals with high Curie temperatures.~\cite{kumar_tunable_2017} 
However, high-throughput density functional theory (DFT) calculations reveal that it is difficult to obtain MAX compounds
with Fe, Co, and Ni,~\cite{ohmer_high-throughput_2019, ohmer_stability_2019} 
 leading to a challenge to realize magnetic MXenes.

Recently, the in-plane ordered MAX (i-MAX) compounds with a formula of (M$_{2/3}$M$'_{{1/3}}$)$_2$AX have been synthesized by substituting ${1/3}$ 
foreign transition metal or rare earth element M$'$  for M in M$_2$AX compounds.~\cite{dahlqvist_prediction_2017,tao_atomically_2019}
Based on theoretical studies for the  i-MAX compounds (Mo$_{2/3}$M$'_{{1/3}}$)$_2$AC (M$'$  = Sc, Y; A = Al, Ga, In, Si,Ge, In),  it was found that the stable conditions for an i-MAX are: (1) the significant difference between the atomic radii for the dopant metal M$'$ and parent M; (2)  small A atoms.~\cite{dahlqvist_origin_2018} 
For the synthesized Cr-bases i-MAX compounds,  DFT calculations suggest (Cr$_{2/3}$Zr$_{{1/3}}$)$_2$AlC is stable in an anti-ferromagnetic spin configuration,~\cite{chen_theoretical_2018} 
while  (Cr$_{2/3}$M$'_{1/3}$)$_2$AlC (M$'$= Sc and Y) are probably stable in a mixed state 
due to small energy difference between the anti-ferromagnetic and ferromagnetic phases.~\cite{lu_theoretical_2017}
This suggests that various magnetic states are possible in the i-MAX compounds, and thus interesting for the corresponding i-MXenes in the monolayer limit. 
On the other hand, the i-MXenes can also be obtained from i-MAX by etching away the A-element atoms, such as the ordered i-MXene W$_{{4/3}}$C and Nb$_{{4/3}}$C.~\cite{meshkian_w-based_2018, zhan_computational_2019}
  
In this work, we performed systematic density functional theory calculations to investigate the magnetic and electronic properties of i-MXenes with a general
chemical formula (M$_{2/3}$M$'_{{1/3}}$)$_2$X, focusing particularly on the cases where M$^\prime$ is magnetic.
It is observed that robust magnetism can be induced, and there exists significant magneto-structural coupling, leading to tunable magnetic ground state and magnetic anisotropy
by strain. Moreover, our calculations suggest that the i-MXenes host fascinating transport properties, such as large Seebeck effect,
antiferromagnetic topological insulators, and spin-gapless semiconductors.

 

\section{Computational method}
\begin{figure}[htp]
\centering

\includegraphics[scale=0.46]{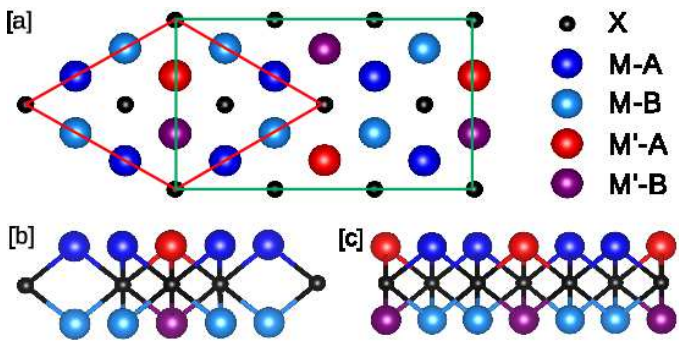}

\label{crysstructure}
\caption{ [a] Top view for i-MXene (M$_{2/3}$M$'_{1/3}$)$_2$X in both hexagonal lattice (red) and rectangular lattice (green). Side views for i-MXene in hexagonal [b] and rectangular [c] lattices.
 The "-A" and "-B" denote the atom above or below the central layer of B in i-MXene.
The original pure MXene crystallizes in a hexagonal lattice. After inducing 1/3 foreign transition metal dopant, the hexagonal symmetry is broken and tilted to the rectangular lattice. 
}
\label{rect-hex-spin}
\end{figure}

To maintain reasonable computational effort, we considered 11 known nonmagnetic MXene as the parent compounds, namely, Sc$_2$C, V$_2$C,  Mo$_2$C, Nb$_2$C, Ta$_2$C, Ti$_2$C, Zr$_2$C, Hf$_2$C, Ti$_2$N, Zr$_2$N, and Hf$_2$N.~\cite{khazaei_insights_2018} 
The dopant element M$'$ is chosen to be one of the transition metal elements except Tc,  resulting in 319 compounds with the general chemical formula (M$_{2/3}$M$'{_{1/3}}$)$_2$X.
As sketched in Fig.~\ref{rect-hex-spin}, the symmetry for the hexagonal MXenes will be lowered after substituting the dopants, leading to a rectangular structure.
In our calculations, both the hexagonal and rectangular structures are considered in order to understand the effect of strain. 
To determine the magnetic ground state, we consider non-magnetic (NM), ferromagnetic (FM), interlayer antiferromagnetic (AFM), and intralayer AFM 
 configurations
for the magnetic M$'$ sublattice (Fig.~S2 in the Supplementary material). 

Our DFT calculations are performed in an automated way using the in-house developed  high-throughput environment,~\cite{zhang_high-throughput_2018, li_high_2018,opahle_high-throughput_2013,opahle_high_2012} 
which is interfaced to Vienna ab initio Simulation Package (VASP)~\cite{kresse_ultrasoft_1999,kresse_efficient_1996} and full-potential local-orbital minimum-basis code (FPLO).~\cite{koepernik_full-potential_1999,opahle_full-potential_1999}
The exchange-correlation functional in the generalized gradient approximation (GGA) is applied, as parameterized by Perdew, Burke, and Ernzerhof (PBE).~\cite{perdew_generalized_1996}
To guarantee good convergence, the plane-wave energy cutoff and $k$-mesh density are set as 500 eV and 60 \AA$^{-1}$, respectively. 
The calculations to obtain optimized structures and various magnetic configurations are carried out using VASP, while the electronic structure and physical properties are obtained using the FPLO and WIEN2k~\cite{noauthor_pdf_nodate} codes, as detailed in our previous work.~\cite{zhang_high-throughput_2018, li_high_2018, gao_high-throughput_2019}

\section{Results and discussions}

\begin{figure}[htp]
\centering
\includegraphics[scale=0.75]{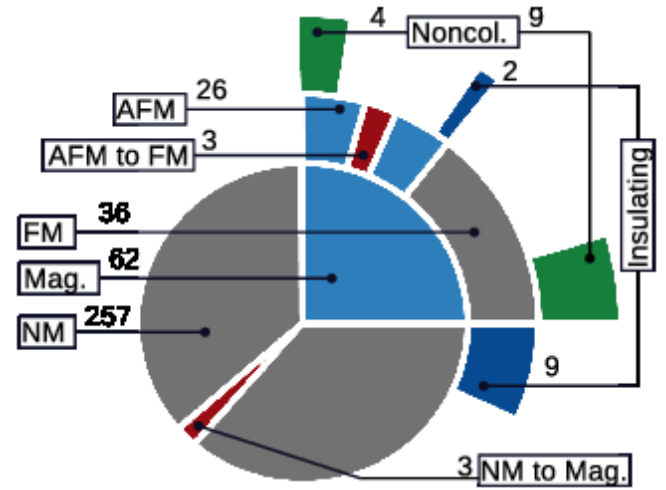}
\caption{Classification of the magnetic ground states for the 319 i-MXene, where "NM to FM" and "AFM to FM" mark the compounds whose ground state 
changes by imposing hexagonal geometry.
}

\label{piechart} 
\end{figure}

As summarized in Fig.~\ref{piechart}, among the 319 i-MXenes in the rectangular geometry, 257 compounds are non-magnetic, and the remaining 62 cases 
are magnetic (with total magnetic moments greater than 0.2$\mu_B$/f.u. in the ferromagnetic spin configuration), where 36 (26) being with ferromagnetic (antiferromagnetic) ground states. 
Furthermore, for the compounds in the imposed hexagonal geometry, the magnetic M$^\prime$ sublattice forms a triangular lattice, which is frustrated and thus can lead
to in-plane noncollinear magnetic structure. 
We found that there are 13 i-MXenes with AFM intralayer exchange coupling between the moments on the M$^\prime$ sites (cf. Table~S1 in Supplementary material), 
indicating possible 2D noncollinear magnetic states. 
Such compounds will be saved for detailed investigation in the future and within the current work we focus only on the collinear magnetic configurations.

Interestingly, comparing the magnetic ground states for i-MXenes in the rectangular and hexagonal geometries, 
there are 6 compounds whose magnetic configurations can be changed (Fig.~\ref{piechart}),  
e.g., 3 NM cases  become FM, and 3 AFM compounds change to FM. 
For instance,  (Zr$_{2/3}$Ti${_{1/3}}$)$_2$N changes from NM (in rectangular lattice) to FM (in hexagonal lattice) states with a total magnetic moment of 0.95 $\mu_B$/f.u..
The same NM$\rightarrow$FM transition occurs for (Ti$_{2/3}$Ru${_{1/3}}$)$_2$C and (Zr$_{2/3}$Cu${_{1/3}}$)$_2$C, with magnetic moments of 0.29 and 0.30 $\mu_B$/f.u., respectively.  
The underlying mechanism can be understood based on the density of states (DOS) of the Zr-5d orbitals as shown in Fig.~S4 in the in the Supplementary material.
Based on the Stoner model, the criterion for a stable ferromagnet is 
\begin{equation}
UN(E_F)>1
\end{equation}
where U is the exchange integral
and $N(E_F)$ is the DOS at Fermi level in the paramagnetic state.
For the rectangular lattice, the Fermi level is located around a peak of DOS.
In the imposed hexagonal structure, the DOS at Fermi level is significantly enhanced from 5.2 states/eV/u.c. to 6.0 states/eV/u.c. (Fig.~S4 in the in the Supplementary material), resulting in the instability of becoming ferromagnetic.
The same behavior is observed also for (Ti$_{2/3}$Ru${_{1/3}}$)$_2$C and (Zr$_{2/3}$Ru${_{1/3}}$)$_2$C (cf. Fig.~S4 in the Supplementary material).

For the compounds with the AFM$\rightarrow$FM transition, such as (Hf$_{2/3}$Fe${_{1/3}}$)$_2$C and (Ti$_{2/3}$V${_{1/3}}$)$_2$C,
the magnetic moments of the Fe (V) atoms change from 1.09 (0.58) $\mu_B$ in the rectangular lattice to as large as 1.79 (0.64) $\mu_B$ in the hexagonal lattice, respectively. 
For (Hf$_{2/3}$Fe${_{1/3}}$)$_2$C, the magnetic moments of Fe atoms are enhanced by 64\%,
which can be attributed to the enhanced exchange splitting (cf. Fig.~S5 in the Supplementary).
It also turns out to be a gapless semiconductor, which will be discussed in detail later.
On the other hand, we have shown the band structures  for (Ti$_{2/3}$V${_{1/3}}$)$_2$C in Fig.~S6 in the in the Supplementary material in both rectangular (Rect.) and hexagonal (Hex.) lattices together with the  anomalous hall conductivity (AHC) for  (Ti$_{2/3}$V${_{1/3}}$)$_2$C in the hexagonal lattice.
In the rectangular lattice, (Ti$_{2/3}$V${_{1/3}}$)$_2$C is AFM with zero AHC.  Obviously, for the FM state in the hexagonal lattice there exists a finite AHC of -160 S/cm at the Fermi level. 
In this sense, strain can be applied to tune the topological transport properties, leading to the piezospintronic effect as discussed in the bulk materials as well.~\cite{boldrin_anomalous_2019,samathrakis_piezospintronic_2019}

It is noted that the lattice deformation from the rectangular to the hexagonal lattices is of marginal magnitude. 
For instance, the strains along the a and b directions for (Ti$_{2/3}$Fe$_{1/3}$)$_2$C are just about 0.14\% and 0.18\%, where the energy difference between rectangular and hexagonal lattices is as small as 1.2 meV/atom. 
Thus, it is straighforward to tailor the i-MXenes from low symmetry rectangular lattice to the hexagonal lattice with higher symmetry. 
On the other hand, it has been previously reported that the piezomagnetic effect can be realized in pure MXene M$_2$C (M = Hf, Nb, Sc, Ta, Ti, V, and Zr) by applying biaxial strain.~\cite{zhao_manipulation_2014} 
In this regard, we suspect that the piezomagnetic effect is  dramatically enhanced in i-MXene as manifested by the five aforementioned cases.
As the hexagonal lattice with high symmetry is more interesting and easy to obtain, we will focus on the physical properties of such systems in the remaining part of this work.

 \begin{table*}

\caption{The basic information of the magnetic i-MXene candidates in hexagonal lattice with an out-of-plane MAE larger than 0.5 meV/f.u., including the MAE in unit of  meV/f.u., the magnetic moment per magnetic atom in unit of $\mu_B$, magnetic order, the exchange coupling parameters and the main magnetic atom. 
}\label{mcae22}
\begin{tabular}{lllllllllllll}
\hline
\hline
	Compound & MAE  & Magnetic& Magnetic&Magnetic& $J_\text{inter}$&$J_\text{intra}$&T$_C$\\
	& meV/f.u. & Moment& order& atom& meV& meV&K\\
	\hline
	(Ta$_{2/3}$Fe${_{1/3}}$)$_2$C  & 0.86& 1.82& AFM &Fe &-10.05&0.56  &-   \\
	(Zr$_{2/3}$Fe${_{1/3}}$)$_2$C  &0.74  &1.71&  FM& Fe &10.16&-2.24 & 267.54 \\
	(Hf$_{2/3}$Fe${_{1/3}}$)$_2$C  & 1.39&  1.79&  FM  & Fe  & 33.94 & 0.70 & 893.67 \\
	(Hf$_{2/3}$Cr${_{1/3}}$)$_2$N&  0.76&1.01&  FM &Cr  & 13.06& 0.20 &343.90  \\
	 (Ti$_{2/3}$Hf${_{1/3}}$)$_2$N& 0.71&0.30&  FM & Ti &  7.22 & 0.97&190.11 \\ 
	 \hline
	 \hline
\end{tabular}

\end{table*}

For two-dimensional magnets, according to the Mermin-Wagner theorem, there exists no long range ordering if there is continuous symmetry for the order parameters.~\cite{mermin_absence_1966}
In this regard, to stabilize 2D magnets at finite temperature, magnetocrystalline anisotropy energy (MAE) is essential, which breaks the rotational symmetry of Heisenberg moments.~\cite{haastrup_computational_2018}
The MAEs for the magnetic i-MXenes are evaluated using the force theorem:~\cite{wang_validity_1996} 
\begin{equation}
\text{MAE}=\sum_{i\in occ.}(\epsilon_{i}^{[100]}-\epsilon_{i}^{[001]})
\end{equation}
where  $\epsilon_{i}^{[001]}$ and $\epsilon_{i}^{[100]}$ denote the energy eigenvalues of the $i-$th band for the magnetization  along [001] and [100] directions, respectively.

Table~\ref{mcae22} lists the 5 i-MXenes with a MAE larger than 0.5 meV/f.u. among all the magnetic i-MXenes (Table~S2 in Supplementary). 
It is noticed that (Hf$_{2/3}$Fe${_{1/3}}$)$_2$C has the largest MAE of 1.39 meV/f.u., favoring out-of-plane magnetization direction.
In addition, the MAEs of (Zr$_{2/3}$Fe${_{1/3}}$)$_2$C and (Ti$_{2/3}$Fe${_{1/3}}$)$_2$C are 0.74 meV/f.u. and 0.03 meV/f.u., respectively.
Such a trend of increasing MAE as X varying from Ti, Zr and Hf can be attributed to the variation of the strength of the atomic spin-orbit coupling (SOC) as 11.2 meV (Ti), 42.1 meV (Zr) and 126.5 meV (Hf).~\cite{dai_effects_2008,martin_table_1971}
Furthermore, the enhanced MAE of (Zr$_{2/3}$Fe${_{1/3}}$)$_2$C  and  (Hf$_{2/3}$Fe${_{1/3}}$)$_2$C is originated from the trigonal crystal fields which lead to strongly SOC coupled bands around the Fermi energy, as manifested by the orbital projected band structures (Fig.~S9 in the supplementary material).
As shown in the orbital projected band structures (Fig.~S9 in the supplementary material), for (Hf$_{2/3}$Fe${_{1/3}}$)$_2$C, the Fe-3d$_{xy}$  and Fe-3d$_{x^2-y^2}$ bands are split by 73 meV along the K-$\Gamma$ section.
While for (Zr$_{2/3}$Fe${_{1/3}}$)$_2$C, the Fe-3d$_{xy}$  and Fe-3d$_{x^2-y^2}$ bands are split by 40 meV along the M-K section.
As the d$_{xy}$  and d$_{x^2-y^2}$ orbitals are strongly coupled by SOC, such a special electronic structure leads to the large MAEs. 
Such a mechanism is in good correspondence to the giant MAE realized in artificial Fe atoms adsorbed on III-V nitride thin films with local trigonal symmetry as well.~\cite{yu_giant_2018}

To obtain the Curie temperature, we took (Hf$_{2/3}$Fe${_{1/3}}$)$_2$C as an example and evaluated the exchange parameters between magnetic Fe atoms by mapping the DFT total energies to the Heisenberg model:
 \begin{equation}
 H=-\frac{1}{2}\sum_{i\neq j}J_{ij}\mathbf{S}_i\cdot\mathbf{S}_j
 \end{equation}
 where $J_{ij}$ is the exchange parameter for the local moments on the $i$ and $j$ sites, $\mathbf{S}_{i/j}$ marks the on-site spin operator. 
Considering three magnetic configurations, $i.e.$, FM, AFM-$\alpha$, and AFM-$\beta$ (cf. Fig.~S2 for spin configurations AFM-$\beta$ and AFM-$\alpha$), the energy differences can be formulated in terms of the interlayer ($J_\text{inter}$) and intralayer ($J_\text{intra}$) exchange parameters (cf. Fig.~S3 in the supplementary material  for $J_\text{intra}$ and $J_\text{inter}$):
 \begin{equation}
E_{FM}-E_{AFM-\alpha}=-J_\text{inter}S^2,
 \end{equation}
  \begin{equation}
E_{FM}-E_{AFM-\beta}=-8J_\text{intra}S^2.
 \end{equation}

Here we use the square of local spin moment length to represent the square of the spin operator. 
The resulting exchange coupling parameters for (Hf$_{2/3}$Fe${_{1/3}}$)$_2$C are: J$_\text{inter}$= 33.94 meV and J$_\text{intra}$= 0.70 meV. 
Similarly, we obtained the exchange coupling parameters for the other ferromagnetic cases such as (Hf$_{2/3}$Cr${_{1/3}}$)$_2$N, (Ta$_{2/3}$Fe${_{1/3}}$)$_2$C,  (Ti$_{2/3}$Hf${_{1/3}}$)$_2$N, and  (Zr$_{2/3}$Fe${_{1/3}}$)$_2$C (Table~\ref{mcae22}). Obviously, the value of interlayer exchange coupling is much larger than that of the intralayer coupling for all the listed compounds. Due to the dramatic large magnetic  anisotropy, we can use the 2D Ising model to estimate the Curie temperature~\cite{onsager_crystal_1944, gibertini_magnetic_2019}
  \begin{equation}
T_{C}=\frac{2J_\text{inter}}{k_Bln(1+\sqrt{2})}.
 \end{equation}
with the results listed in Table~\ref{mcae22}. 
Surprisingly, the Curie temperatures for (Hf$_{2/3}$Fe${_{1/3}}$)$_2$C and (Hf$_{2/3}$Cr${_{1/3}}$)$_2$N are even above the room temperature. 
 It is noted that for the recently synthesized 2D magnet CrI$_3$ with a Curie temperature of 45 K, its MAE is about 1.71 meV/f.u.~\cite{huang_layer-dependent_2017, huang_electrical_2018, mcguire_coupling_2015, haastrup_computational_2018} 
Obviously, the out-of-plane MAEs of (Hf$_{2/3}$Fe${_{1/3}}$)$_2$C and (Hf$_{2/3}$Cr${_{1/3}}$)$_2$N are almost the same as that of CrI$_3$.
Futhermore, for CrI$_3$  the interlayer and intralayer exchange  parameters are 11.64 meV and 2.37 meV.~\cite{besbes_microscopic_2019}  So, the exchange parameters for (Hf$_{2/3}$Fe${_{1/3}}$)$_2$C and (Hf$_{2/3}$Cr${_{1/3}}$)$_2$N are larger than that of  CrI$_3$.   
In this point of view, we suspect the i-MXene (Hf$_{2/3}$Fe${_{1/3}}$)$_2$C and (Hf$_{2/3}$Cr${_{1/3}}$)$_2$N being promising 2D magnets with high Curie temperature.

\section{Electronic properties}

\subsection{Thermoelectric properties of semiconductors}

\begin{figure}[htp]
\centering
[a]\includegraphics[scale=0.32]{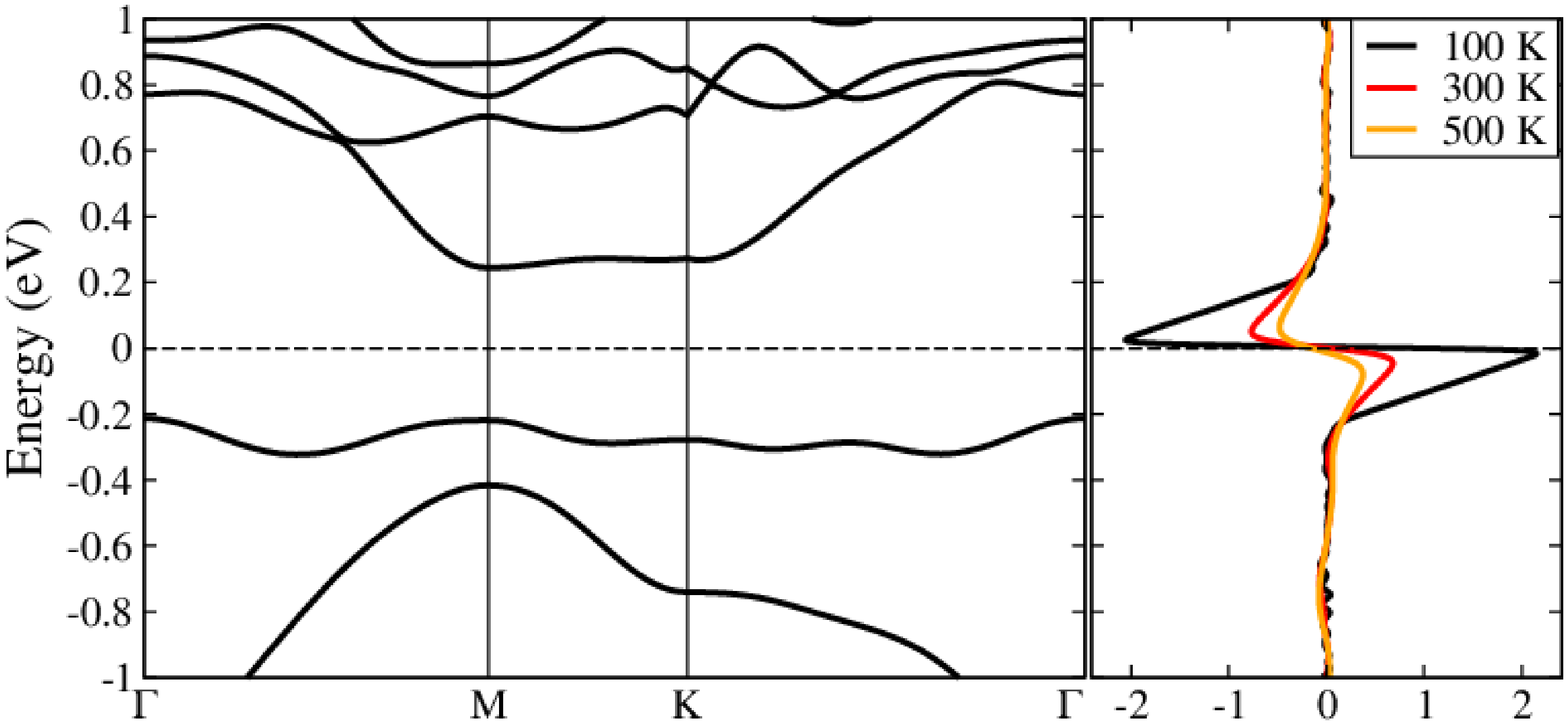}
[b]\includegraphics[scale=0.32]{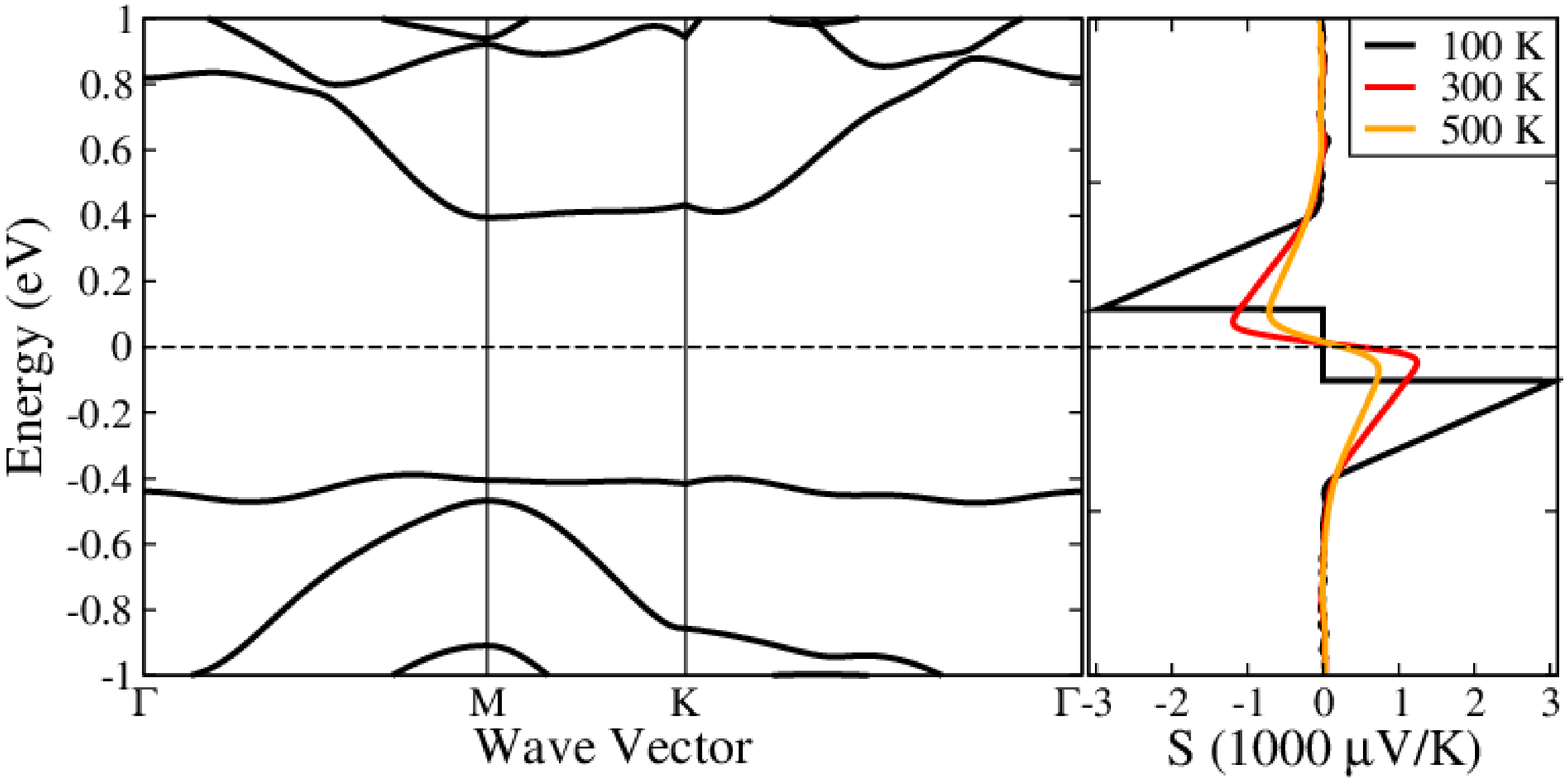}
\caption{Band structures and Seebeck coefficients for (Sc$_{2/3}$Cd${_{1/3}}$)$_2$C [a] and (Sc$_{2/3}$Hg${_{1/3}}$)$_2$C [b]. The horizontal dashed lines denote the Fermi level.
}
\label{seebeck}
\end{figure}

Previously, it has been reported that the 2D semiconductors have very large thermoelectric power.~\cite{zhao_ultrahigh_2016, shafique_thermoelectric_2017} 
For instance, SnSe monolayers are regarded as promising thermoelectric material, $e.g.$, the Seebeck coefficient is increased from 160 $\mu_V$/K to 300 $\mu_V$/K when temperature is increased from 300 K to 700 K.~\cite{zhao_ultrahigh_2016}
Although most i-MXenes tend to be metallic, there exist nine nonmagnetic semiconductors $i.e.$, (Sc$_{2/3}$X${_{1/3}}$)$_2$C (X = Au, Cu, Ir, Ni, Zn, Cd, Hg), (Hf$_{2/3}$Ir${_{1/3}}$)$_2$C and (Ti$_{2/3}$Au${_{1/3}}$)$_2$C.
 Among them,  (Sc$_{2/3}$Cd${_{1/3}}$)$_2$C and (Sc$_{2/3}$Hg${_{1/3}}$)$_2$C are large gap semiconductors 
 (with finite band gap larger than 0.4 eV and less than 1.0 eV), which can be good candidate thermoelectric materials.
The band structures of these two semiconductors in the hexagonal lattice are displayed in Fig.~\ref{seebeck} together with the Seebeck coefficients as a function of chemical potential. 
The band gaps of  (Sc$_{2/3}$Cd${_{1/3}}$)$_2$C and (Sc$_{2/3}$Hg${_{1/3}}$)$_2$C are as large as 0.41 and 0.79 eV, respectively.
Furthermore, the valence band maximum (VBM) and conduction band  minimum (CBM) are flat around the Fermi level. Such a behavior implies that both compounds 
have the potential to realize large Seebeck effect. 
To confirm this, the Seebeck coefficients have been evaluated for such two compounds, as shown in Fig.~\ref{seebeck}. 
The peak of Seebeck coefficients around Fermi level for (Sc$_{2/3}$Cd${_{1/3}}$)$_2$C ((Sc$_{2/3}$Hg${_{1/3}}$)$_2$C) are remarkably as high as 2100 (3000) $\mu$V/K at 100 K, 820 (1200) $\mu$V/K at 300 K, and 520 (700) $\mu$V/K at 500 K. 
The enhanced Seebeck coefficients can be attributed to the large derivative of the DOS with respect to the energy and hence the flat bands above and below the Fermi energy.
It is noted that the density of states for these two systems are very comparable for the hexagonal and rectangular lattices, 
leading to negligible changes in the Seebeck coefficients (cf. Fig.~S7 and S8 in Supplementary). 
Therefore, we suspect that (Sc$_{2/3}$Cd${_{1/3}}$)$_2$C and (Sc$_{2/3}$Hg${_{1/3}}$)$_2$C are good candidates as 2D thermoelectric materials.

\subsection{Topological properties}
\begin{figure}[htp]
\centering
[a]\includegraphics[scale=0.30]{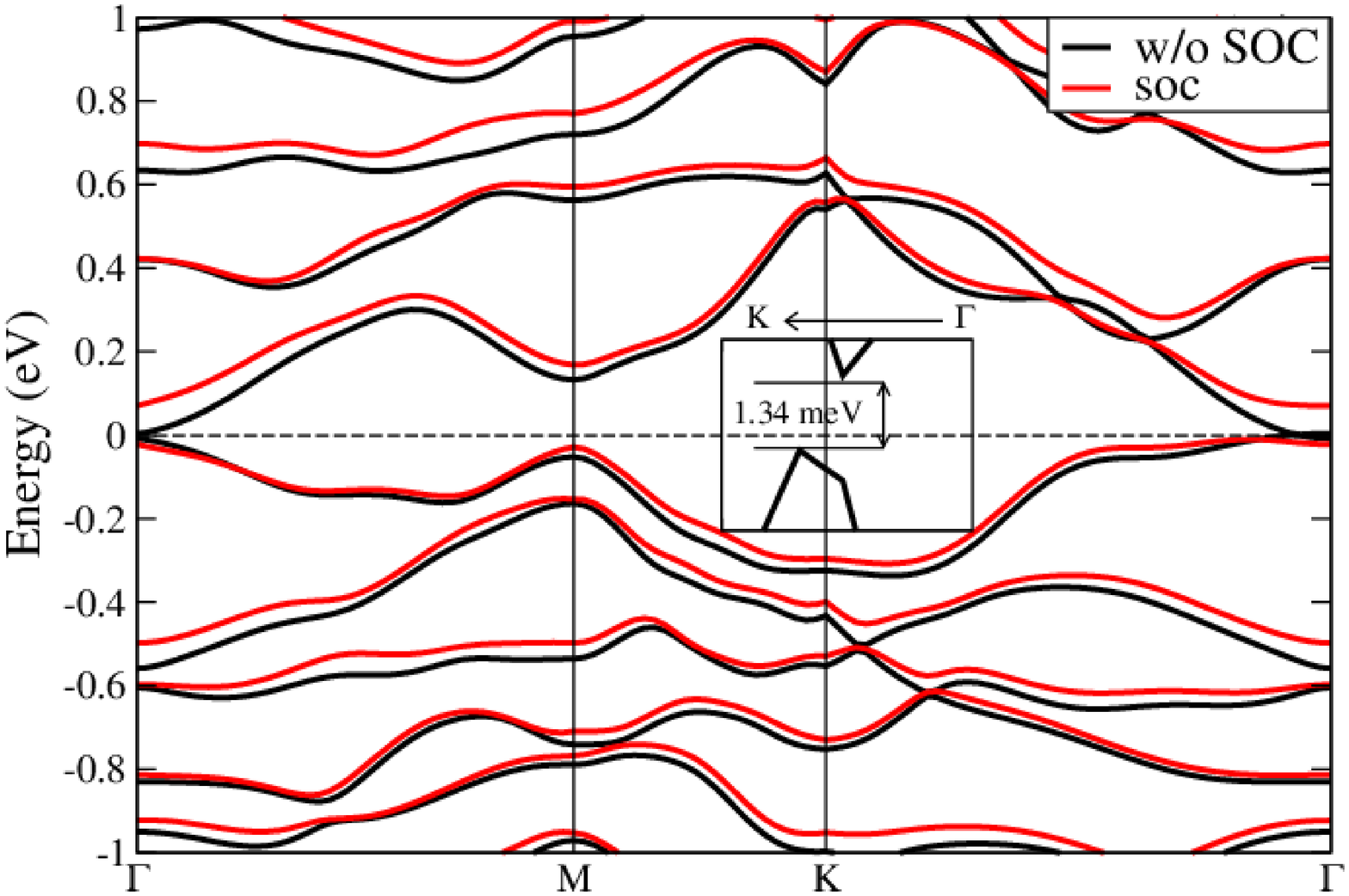}
[b]\includegraphics[scale=0.32]{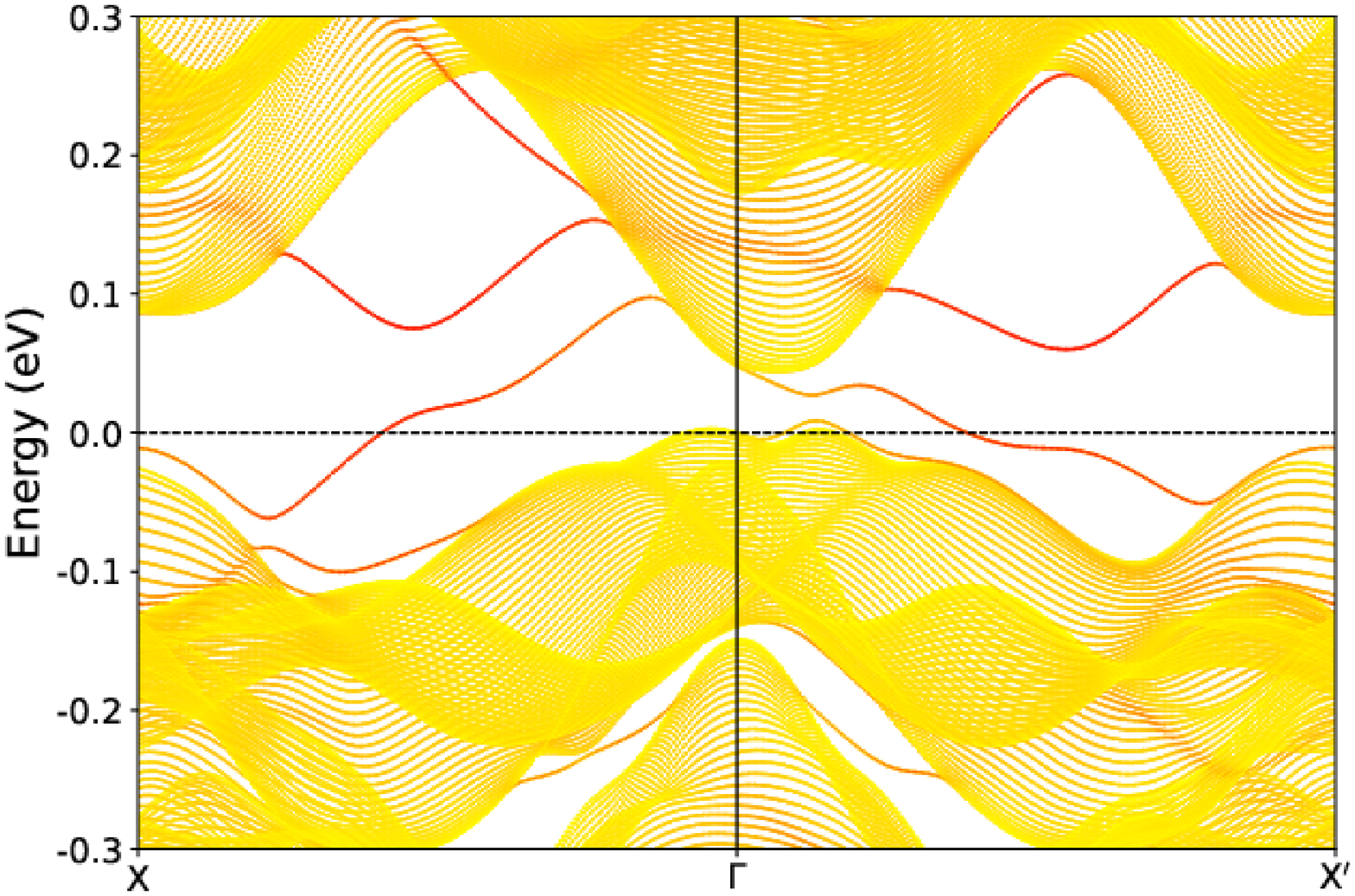}
\caption{Band structures [a] and edge states [b] for AFM topological insulator (Ta$_{2/3}$Fe${_{1/3}}$)$_2$C. When SOC is turned off, the band structures of spin up channel overlap with that of the spin down channel. So we only show band structure in one spin channel. In both [a] and [b], the dashed line denote the Fermi level.}
\label{topo}
\end{figure}

As discussed above, most i-MXenes are metallic due to the partially filled $d$-shells of the transition metal atoms.
However, motivated by the reported nontrivial topological state in some functionalized MXenes,~\cite{si_large-gap_2016}
we studied also the topological nature of the  i-MXene (Ta$_{2/3}$Fe${_{1/3}}$)$_2$C  with a tiny band gap of 1.34 meV.
Surprisingly, (Ta$_{2/3}$Fe${_{1/3}}$)$_2$C with a collinear AFM ground state hosts a nontrivial state, as shown in the bulk and surface electronic structures in Fig.~\ref{topo}. 
Without considering spin-orbit coupling (SOC), the VBM and CBM almost cross each other at the Fermi energy. 
According to the fat band analysis (Fig.~S9 and S10 in Supplementary), these two bands are mainly contributed by the Fe-3d$_{\text yz}$ and Fe-3d$_{\text zx}$ orbital character, which is strongly coupled by SOC.
Turning on SOC, a remarkable indirect gap of 75.0 meV is opened, 
leading to band inversion and thus the occurrence of the topological nontrivial state. 
This is clearly confirmed by our explicit calculation of the edge states of  the corresponding 1D ribbons with a width of 70 units, as shown in Fig.~\ref{topo}(b).
Furthermore, the nontrivial edge state is protected by the combination of the time reversal 
($\Lt$) and space inversion ($\Lp$) symmetries ($\Lt\Lp$), as demonstrated for CuMnAs in Ref.~[\onlinecite{smejkal_electric_2017}].

\subsection{Spin-gapless Semoconductors}
\begin{figure}[htp]
\centering
[a]\includegraphics[scale=0.32]{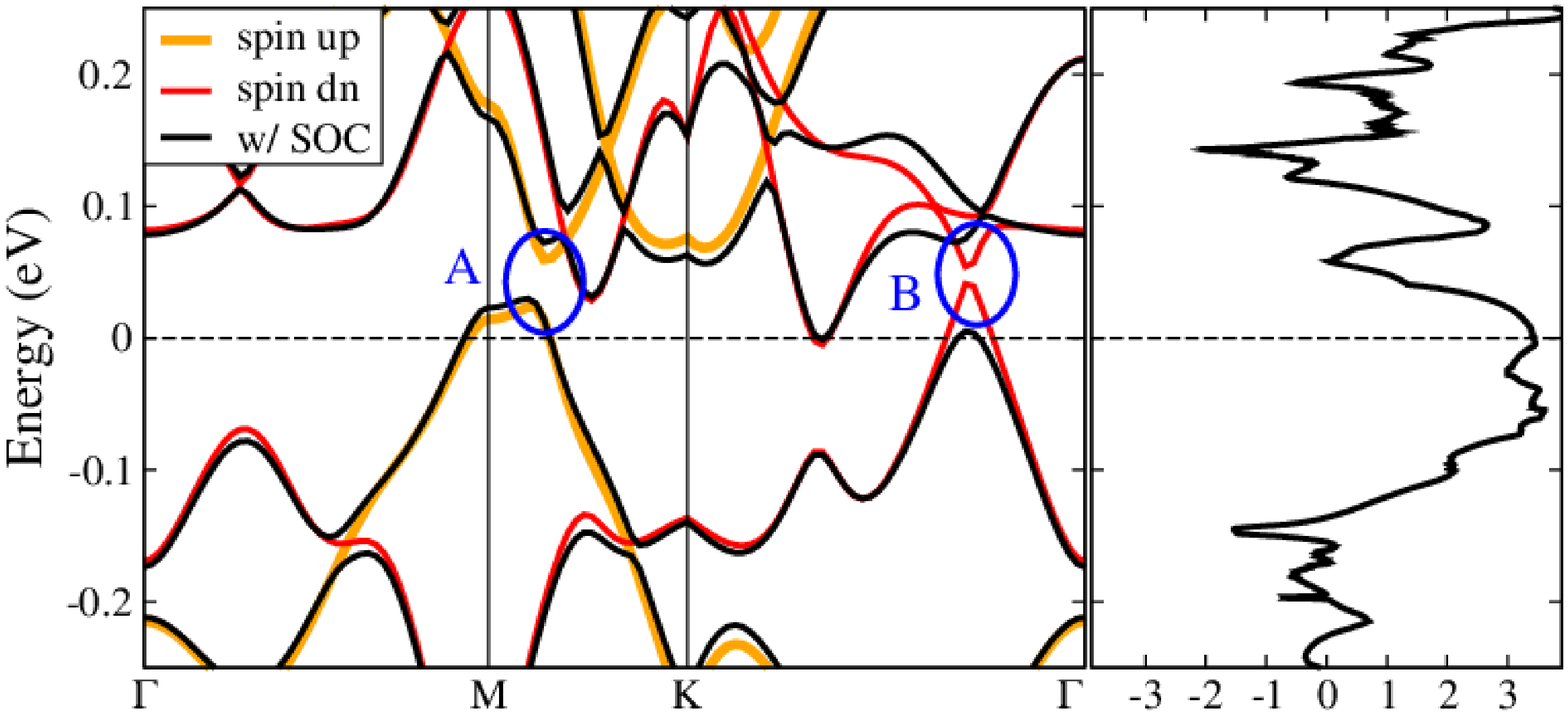}

[b]\includegraphics[scale=0.32]{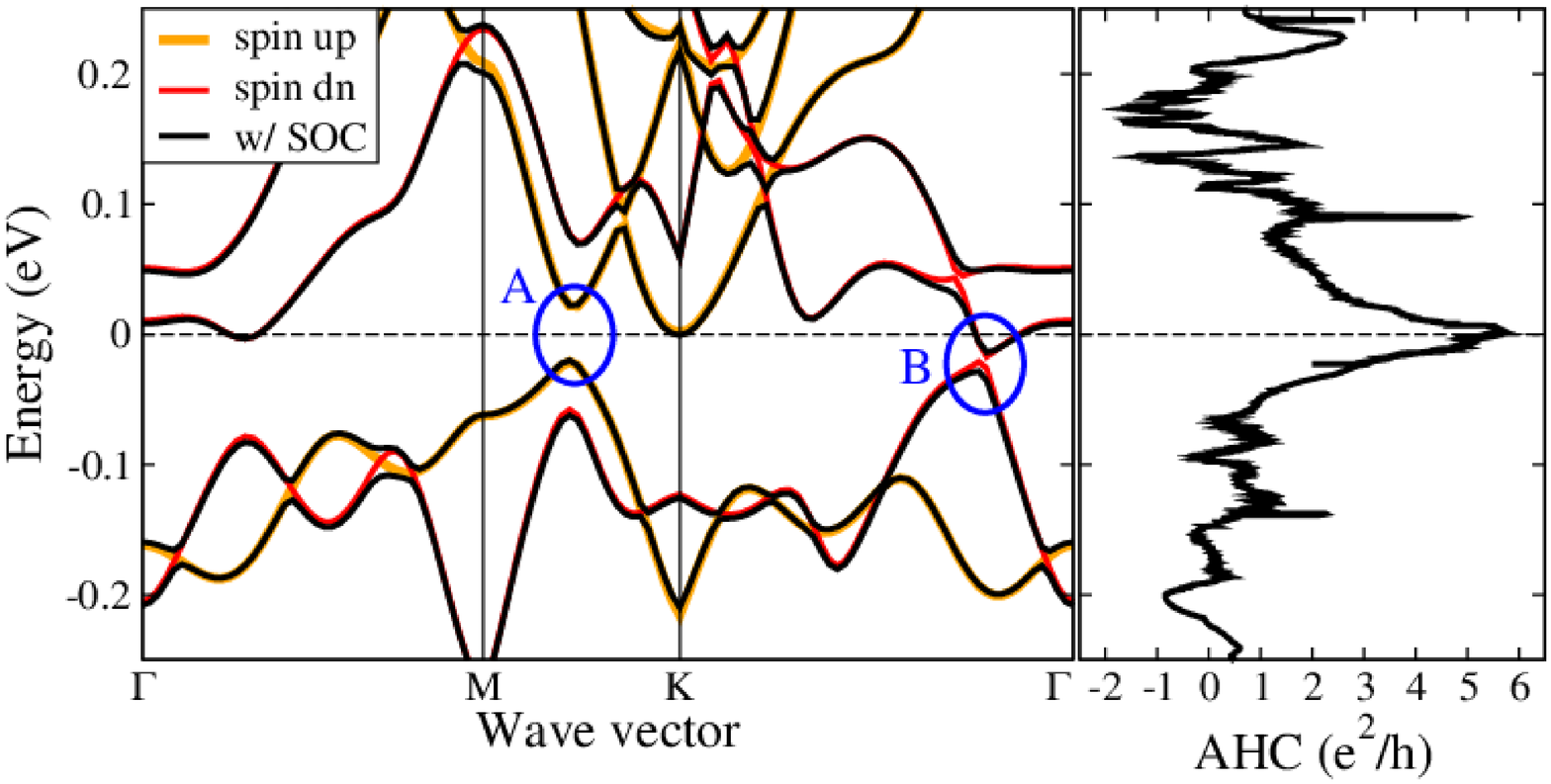}

\caption{The band structures (left) and anomalous Hall conductivity (right) for (Hf$_{2/3}$Fe${_{1/3}}$)$_2$C [a] and (Zr$_{2/3}$Fe${_{1/3}}$)$_2$C [b]. 
The horizontal dashed lines denote the Fermi energies.
}
\label{sgs}
\end{figure}

It is reported that MXene Ti$_2$C can be a spin-gapless semiconductor (SGS) under 2\% strain.~\cite{gao_monolayer_2016} 
We found two SGS candidates among i-MXenes in the hexagonal geometry, {\it e.g.}, (Hf$_{2/3}$Fe${_{1/3}}$)$_2$C and (Zr$_{2/3}$Fe${_{1/3}}$)$_2$C with total magnetic moments of 1.79 and 0.74 $\mu_B$/f.u., respectively.
Such two SGSs are not within the classification of four types SGSs defined in our previous work,~\cite{gao_high-throughput_2019}  because the VBM and CBM 
have both spin characters as shown in Fig.~\ref{sgs}.
For such tow SGSs at Fermi level,  in the spin up channel the CBM and VBM touches each other directly between $K-\Gamma$  (in A region), while in the spin down channel there is only a small gap between for the CBM and VBM between $K-M$ (in region B). That is the VBM and CBM of spin up and down channels can be roughly seen  at the same energy level.
When SOC is switched on, the touching bands open a local band gap, suggesting a topologically nontrivial state. 
This can be manifested by explicit evaluation of the anomalous Hall conductivity. 
For instance, it is observed that the anomalous Hall conductivity is finite around the Fermi energy, 
which is almost quantized to 3 and 5 $e^2/h$ with tiny band gaps of 3.8 meV and 0.3 meV for  (Hf$_{2/3}$Fe${_{1/3}}$)$_2$C and (Zr$_{2/3}$Fe${_{1/3}}$)$_2$C, respectively. 
That is, such i-MXenes are Chern insulators with a nontrivial Chern number of 3 and 5, though the band gaps are small.

\section{Conclusion}
In conclusion, we have done a systematic study on the magnetic and electronic properties of i-MXene compounds, which provide an interesting playground 
for multifunctional applications. 
The magnetic ground states for i-MXene are investigated for both rectangular and hexagonal lattices, where slight strain can be applied to tune the 
magnetic ground state.
Due to the underlying crystal fields, the magnetocrystalline anisotropy of i-MXene can be significantly enhanced, i.e., 
there are 7 systems with a magnetic crystalline anisotropy energy larger than 0.5 meV/f.u..
Furthermore, investigation on the electronic properties reveals that i-MXene can host fascinating transport properties, including significant thermoelectric effects,
antiferromagnetic topological insulator state, and spin gapless semiconductors.
Our calculations suggest that i-MXene is a class of 2D materials which are promising for future applications, calling for further experimental exploration.

\section*{ACKNOWLEDGMENTS}

Qiang Gao thanks the financial support from the China Scholarship Council. The authors gratefully acknowledge the computational time on the Lichtenberg High Performance Supercomputer.


\footnotesize{

\bibliographystyle{plain}
\bibliography{reference}

}

\end{document}